\documentclass[preprint,aps,showkeys,showpacs]{revtex4-1}   
\usepackage{epsfig}
\usepackage{xcolor}   
\usepackage{amsmath}

\begin{document}

\title{Quantum droplets in three-dimensional Bose-Einstein condensates}

\author{Sherzod R. Otajonov}

\address{Physical-Technical Institute of the Uzbek Academy of Sciences,\\
Chingiz Aytmatov Str. 2-B, Tashkent, 100084, Uzbekistan}

\begin{abstract}
The properties of 3D Bose-Einstein condensate have been studied with variational and numerical methods. In the variational approach, we use the super-Gaussian trial function, and it is demonstrated that this trial function gives a good approach for the descriptions of the quantum droplets. The analytical equations for the variational parameters are obtained. The frequency of small oscillations of quantum droplets near the equilibrium position is estimated.   It is also found that periodic modulation of the coupling constants leads to the resonance oscillations of the quantum droplets parameters or emission of waves depending on the amplitude of the modulations. The predictions are supported by direct numerical simulation of governing equation. 
\end{abstract}


\maketitle

\section{Introduction}
\label{sec:intro}
From the mean-field theory, a Bose-Einstein condensate (BEC) in two- and three-dimensions is expected to collapse with the attraction between atoms. There are several methods to stabilize the BECs such as an application of external traps, an account of the three-body interaction term or dipolar interactions, and periodic variation of scattering parameters~\cite{Donley2001, Cornish2000, Roberts2001, Koch2008, Abdullaev2003}. 
In 2015, Petrov suggested that quantum fluctuations can stabilize localized waves in a two-component BEC, consist of two different kinds of atoms or the same atomic species in different states~\cite{Petrov2015}. 
Quantum fluctuations arise from the influence of thermal cloud (non-condensed atoms) on a coherent Bose gas. This influence is taken into account by the first-order correction to the condensate energy, known as the Lee-Huang-Yang (LHY) term~\cite{LHY}.

In a single-component BEC the LHY term is negligible with respect to the mean-field term.
In a two-component BEC, the scattering parameters can be changed via the Feshbach resonance technique~\cite{Chin2010} such that there are intra-species (interaction between the same atomic species) repulsion and inter-species (interaction between different atoms) attraction. Then, in absence of quantum fluctuations, such a BEC collapses. This residual attraction can be balanced by repulsive interaction due to quantum fluctuations. As a result of this balance, the density distribution of a BEC takes a localized profile. This state is associated with a quantum droplet (QD) because it has liquid-like properties. 

Different aspects of QDs have been studied in a large number of papers. The dimensional reduction from 3D to 2D and 1D are reported in Ref.~\cite{Petrov2016}. Dynamics of 1D QDs was studied in ~\cite{Astrakharchik2018, Otajonov2019}. Generation of the QDs through modulational instability in the 1D binary condensate is investigated in~\cite{Mithun2020}. Two-dimensional QDs and vortices have been considered in~\cite{Li2018, Otajonov2020}.
In 3D cases, QDs and vortices have been studied in Ref.~\cite{Kartashov2018} by means of the numerical methods, the stability regions of the vortex QDs are found for vortices $S=1$ and $2$, where $S$ is a topological charge. It is also shown that in the parameter domains all the hidden vortex states are completely unstable. 
Recently the existence of Lee-Huang-Yang fluid was shown in a bosonic mixture where scattering parameters and density of the components are tuned such that the residual mean-field interactions entirely cancel out as a result system is governed only by quantum fluctuations~\cite{Jorgensen2018}.
The formation of quantum balls and the role of three-body interaction in binary BEC have been reported in Ref.~\cite{Gautam2019}.

In dipolar BECs, the parameter of dipolar interaction can be varied independently on the strength of quantum fluctuations. Therefore, QDs can be formed even in a single-component dipolar BEC~\cite{Smith2021,Bisset2021}. Quantum droplets can also be formed in other physical systems such as spin-orbit~\cite{Li2017,Tononi2019,Baena2020}, and Bose-Fermi~\cite{Adhikari2018,Rakshit2019} BECs.
An experimental realization of QDs has been also subjected to extensive research~\cite{Ferioli2019, Barbut2016, Edler2017, Cabrera2018, Semeghini2018, Skov2021}.

The Lee-Huang-Yang correction to the condensate energy is repulsive, $\sim n^{5/2}$ in 3D, where $n$ is the condensate density. In 2D, the LHY correction $\sim n^2 \log(n/\sqrt{e})$ and the sign of the LHY term changes depending on the value of n. In 1D, the LHY correction is attractive, $\sim n^{3/2}$. Therefore, the behaviour and dynamics of localized matter waves in a presence of quantum fluctuations depend strongly on dimensionality of a system. Studying the properties of 3D QDs is particularly interesting because it is close to real experiments. 

An important property of QDs is a saturation of the peak density for large number $N$ of atoms. This means that a description of the BEC density of a QD tends to a flat-top profile. A description of such profile with the Gaussian function only works for small $N$. In Refs.~\cite{Otajonov2019, Otajonov2020} it is demonstrated that the super-Gaussian function characterizes well 1D and 2D QDs.
Thus, the main purpose of this paper is to develop a variational approach (VA) based on the super-Gaussian trial function for 3D QDs. This approach provides a good estimate of the static and dynamic parameters of QDs. It also allows us to find an eigenfrequency for which droplets are very sensitive. 

The paper is structured as follows: Sec.~\ref{sec:model} is devoted to the description of the model and variational approximation, in Sec.~\ref{sec:modulation} the dynamics of QDs under the periodic modulations of parameters are presented.
In Sec.~\ref{sec:simulation} numerical simulations and estimations of realistic parameters are presented. In the last Sec.~\ref{sec:conc}, we summarize our findings.

\section{The model and variational approximation }
\label{sec:model}
Let us consider a two-component Bose-Einstein condensate in 3D. 
The mass of atoms as well as the number of atoms in each component is set to be equal. In this case, the system is described by a single  Gross-Pitaevskii equation (GPE)~\cite{Petrov2015}.
\begin{eqnarray}
  & i \hbar \Psi_T+ \cfrac{\hbar^2}{2 m_0} \nabla ^{2} \Psi + \cfrac{|\delta g|}{2} |\Psi |^{2} \Psi 
-\cfrac{4 m_0^{3/2} g^{5/2}}{3 \pi^2 \hbar^3} |\Psi|^{3} \Psi =0,
\label{dimgpe}
\end{eqnarray}
where $\Psi=\Psi(R,T)$ is the BEC wave function, $m_0$ is the atomic mass, $\nabla^2$  is the three-dimensional Laplacian, 
$T$ is the time, $\Psi_T \equiv \partial \Psi / \partial T$, $\delta g=g_{12}+g<0$ is the modified coupling constant with $g_{11}=g_{22}=g=4 \pi \hbar^2 a/m_0>0$, 
and $g_{12}=4 \pi \hbar^2 a_{12}/m_0<0$ are the intra- and inter-species coupling constants, respectively.
For simplicity, we do not include the effect of gravity in our model, experimentally such a BEC can be realized in free-fall experiments. Another alternative way is to generalize these results with time-averaged optical potentials. 
We consider localized states of condensate with a finite number of atoms in an infinite system, $|\vec{R}|<\infty$.

By using new variables $t=T / t_s$, $(x,y,z)=(X,Y,Z)/ r_s$, and
$\psi=\Psi / \psi_s$\,, we reduce the Eq.~(\ref{dimgpe}) in the following dimensionless form: 
\begin{eqnarray}
      i \psi_{t}+ \cfrac{ 1 }{ 2} \nabla ^{2} \psi + \alpha |\psi |^{2} \psi - \beta |\psi|^{3} \psi =0\,.
\label{gpe}
\end{eqnarray}
where $\psi_t \equiv \partial \psi / \partial t $,
and the scale parameters are 
$$t_s=\cfrac{128 \alpha^3 m_0^3 g^5}{9 \pi^4 \beta^2 \hbar^5 |\delta g|^3 }\,, \quad
r_s=\cfrac{8\sqrt{2} m_0 \alpha^{3/2} g^{5/2} }{3 \pi^2 \hbar^2 \beta |\delta g|^{3/2}}\,,$$  
$$\psi_s=\cfrac{3 \pi^2 \hbar^3 \, \beta \, |\delta g|}{8 \alpha m_0^{3/2} g^{5/2} }\, .$$
The parameters $\alpha$ and $\beta$ can be chosen arbitrarily and in the dimensionless equation, they represent the strength of the mean-field and quantum fluctuation interaction terms, respectively. We keep these parameters so that if one wants to extend this study for the cases when either effects two-body interaction or quantum fluctuations are ignored. By choosing the $g_{12}=g$ the two-body interaction term can be ignored ($\alpha=\delta g=0$) from the Eq.~(\ref{gpe}). In this case, the repulsive quantum fluctuation can be balanced with an external trap. Such a system is known as LHY fluid and was first studied theoretically in Ref.~\cite{Jorgensen2018} and experimentally in Ref.~\cite{Skov2021}.

In our analysis, we consider a spherically symmetric system, which reduces Eq.~(\ref{gpe}) into a quasi-one-dimensional equation that only depends on the radial coordinate.
We consider only spherical symmetric perturbations because this is the basic type of modulations. We assume such types of modulations easier to induce in experiments rather than high order (multi-pole) modulations. The form of the Laplacian in spherical coordinate is given by
$\nabla^{2}=\cfrac{  \partial^{2} }{ \partial r^{2} }+ \cfrac{ 2 }{ r} \cfrac{\partial}{\partial r} $\,.

The Lagrangian density of Eq.~(\ref{gpe}) is
\begin{equation}
    \mathcal{L} = \cfrac{i }{ 2} (\psi_{t}^{*} \psi - \psi^{*} \psi_{t})
     +\cfrac{1 }{ 2} |\psi_{r}|^{2}- \cfrac{\alpha }{ 2}|\psi|^{4}
     + \cfrac{2 \beta }{ 5} |\psi|^{5} \,,
\label{lagrdens}
\end{equation}
where $\psi_r \equiv \partial \psi/ \partial r$. 

We employ the following super-Gaussian trial function:
\begin{equation}
  \psi(r,t)= A \exp\left(- \cfrac{1 }{ 2} \left( \cfrac{r }{ w}\right)^{2m} +ibr^{2} + i\varphi\right),
\label{ansatz}
\end{equation}
where $A(t),w(t),b(t)$ and $\varphi(t)$ are the variational parameters, denoting the amplitude
width, chirp and initial phase, respectively. We assume that, super-Gaussian indices $m$ does not depend on time. The value of $m$ is determined from the parameters of stationary solution $m=m_s$, see Eqs.~(\ref{wtLm}). The advantage of choosing this ansatz is that 
it allows to describe the small (bell shape) and as well as large droplet (flat-top shape) states.

The norm
\begin{equation}
     N=4\pi \int \limits_{0}^{\infty}{r^{2}|\psi|^{2} dr} ={4 \over 3 } \pi A^{2} w^{3} \Gamma(1+3M)
\label{norm}
\end{equation}
is a conserved quantity and it is proportional to the number of atoms in the droplets. The parameter $M={1/2m}$  is reduced super-Gaussian indices.
Substituting trial function into Eq.~(\ref{lagrdens}), integrations over space $L = 4\pi \int \limits_{0}^{\infty }{r^2 \mathcal{L} dr}$
and using Eq.~(\ref{norm}) to eliminate $A$ yields the averaged Lagrangian.
\begin{eqnarray}
& \cfrac{L}{N}= \varphi_{t} +\cfrac{ 3 \Gamma(1+5M)}{ 5 \Gamma(1+3M) }w^{2}(2b^{2} +b_{t}) +G(M,w,N),
\nonumber \\
\label{lagr}
\end{eqnarray}
where
\begin{equation}
  G(M,w,N) = \cfrac{3 \Gamma(M+2)}{8 M \Gamma(3M+1) w^{2} } - \cfrac{G_1 \alpha }{w^3}
  +\cfrac{G_2 \beta}{ w^{9/2} } \,,
\label{G}
\end{equation}
\begin{equation}
  G_1 \equiv \cfrac{3 \cdot 2^{-3(M+1)} N  }{ \pi \Gamma(3M+1)}, \quad G_2 \equiv \cfrac{3 \sqrt{3}\, 2^{3M-2} }{5^{3M+1}} \left( \cfrac{N }{\pi \Gamma(3M+1) } \right)^{3/2}.
\label{G012}
\end{equation}
The Euler-Lagrangian equations leads to the following system of equations:
\begin{equation}
  b_{t} = -2{{b}^{2}}-{5 \Gamma(3M+1) \over 6 w \Gamma(5M+1)} {\partial G \over \partial w} \equiv f_{b} \, ,
\label{bt}
\end{equation}
\begin{equation}
 w_{t}=2bw \equiv f_{w}\, , \qquad  \cfrac{\partial L}{\partial M} \equiv {{f}_{m}}=0 \, .
\label{wtLm}
\end{equation}
To find super-Gaussian parameter $m$ we use the same technique as it shown in Ref.~\cite{Otajonov2019, Otajonov2020}, taking
$b = b_{t}=\varphi_{t}=0$ from $f_{b} (w,m,N )=0$ we find $w$,
and substituting it into the $f_{m}(w,m,N)=0$ and root of this equation gives $m$ for fixed parameters $(N,\alpha,\beta)$.
For a given parameters $\alpha$ and $\beta$, there is a threshold of $N_{th}$, where $N<N_{th}$ QDs do not exist. This threshold can be found from the condition of appearance of roots of equation $f_{m}(w,m,N)=0$. In Ref.~\cite{Petrov2015}, for parameters $(\alpha, \beta)=(3,5/2)$ $N_{th}$  is found by solving the Bogoliubov-de Gennes equations the value of the threshold is $N_{th} \approx 18.65$.
It is stated in Ref.~\citep{Ferioli2020} that, this threshold value of $N$ was also confirmed in experiments~\citep{Cabrera2018, Semeghini2018}. That is why we use the same parameters for comparison of VA with the results of Ref.~\citep{Petrov2015, Cabrera2018, Semeghini2018}. In VA we found the threshold value as $N_{th}\simeq19.5$ which is very close to the $18.65$, and shows the validation of the proposed approach.

The equation for the width is obtained by using Eq.~(\ref{bt}) and the former of Eq.~(\ref{wtLm})
\begin{equation}
  w_{tt} = -{5 \Gamma(3M+1) \over 3 \Gamma(5M+1)}
    \cfrac{\partial G}{\partial w}=-{\partial U(w) \over \partial w}\, .
\label{wtt}
\end{equation}
The corresponding effective potential is
\begin{equation}
  U(w) = {5 \Gamma(3M+1) \over 3 \Gamma(5M+1)}\, G \, .
\label{pot}
\end{equation}

\begin{figure}[htbp]
\centerline{ \includegraphics[width=6.6cm]{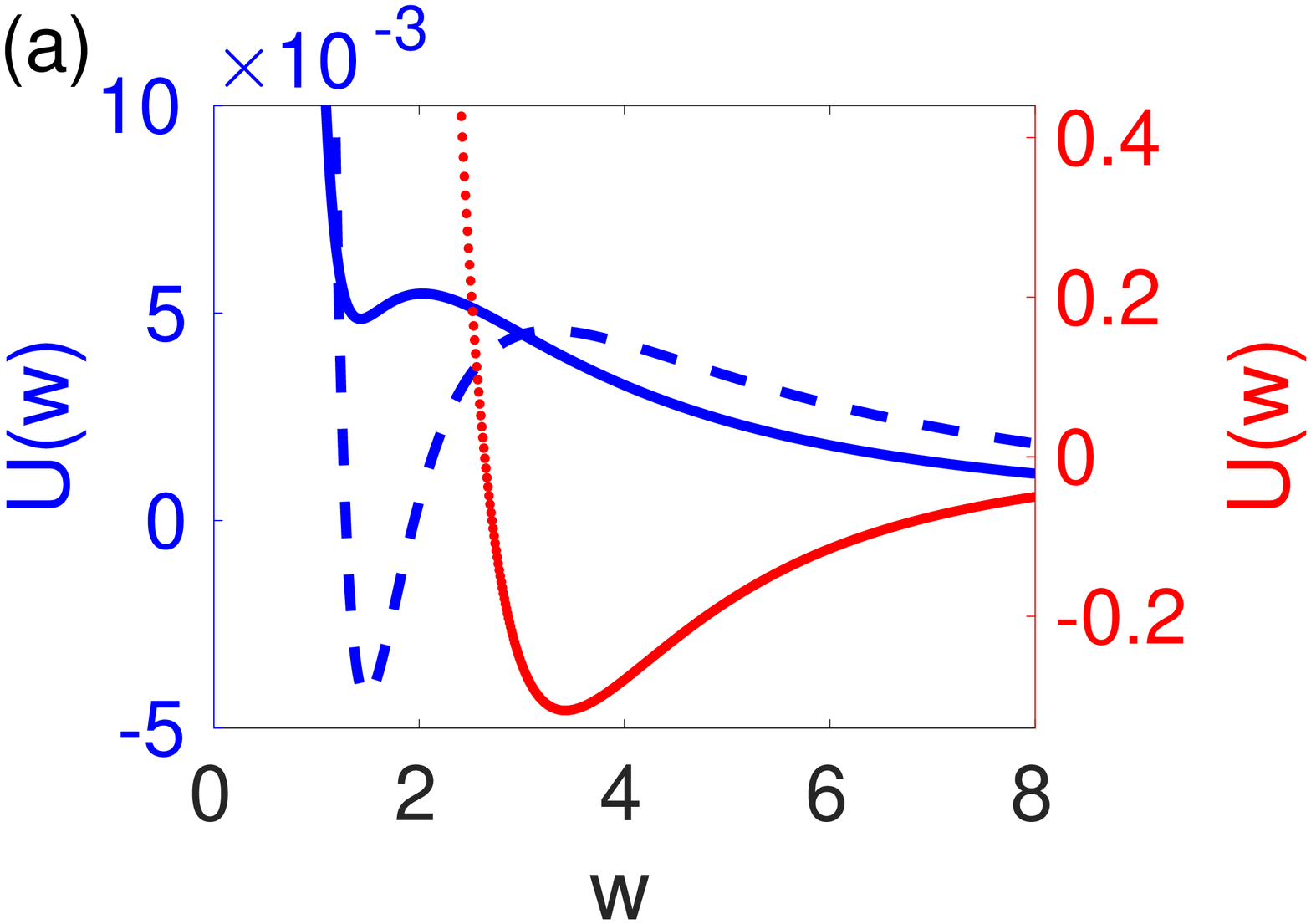} \hskip-0.1cm
\includegraphics[width=6.4cm]{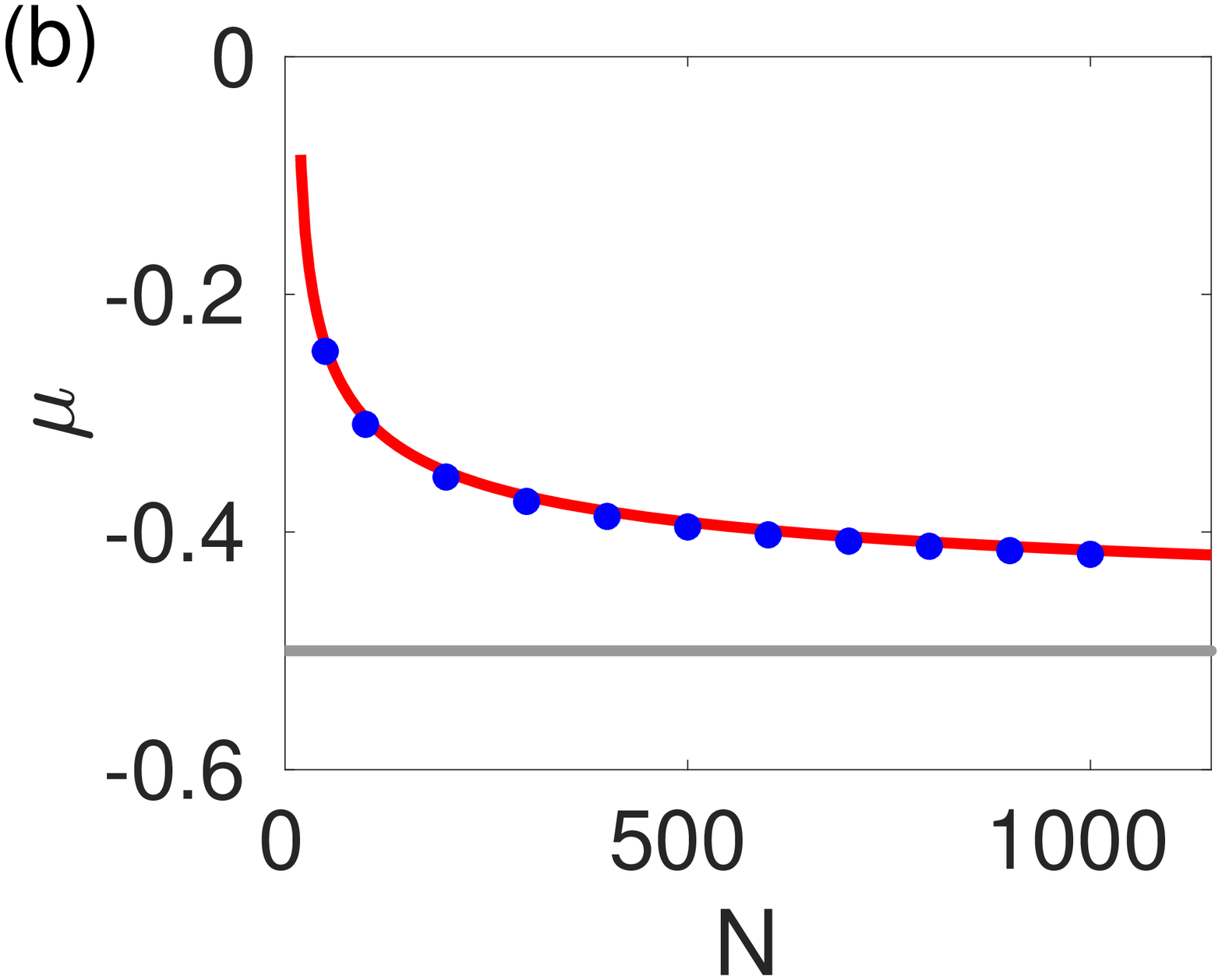} }
\caption{(a) The shape of effective potentials for different values of $N$. The solid line dashed line, and points (right axes) are for 20, 25, and 200, respectively. (b) The chemical potential $\mu$ as a function of norm $N$, for the parameters $(\alpha, \beta)=(3,5/2)$. The red line is from Eq.~(\ref{mu}), points are found from
imaginary time simulations. The gray line represents the Thomas-Fermi limit.
}
\label{fig:muU}
\end{figure}

Typical shapes of the effective potentials are plotted in Fig.\ref{fig:muU}(a) for the values of $N$.
The solid line corresponds to the metastable droplet. In Ref.~\cite{Petrov2015}, by analysing the droplet energy numerically it is shown that such QD is metastable for $18.65 \lesssim$ $N<22.55$. In VA, we found the metastable regions from the $U(w)$ potential curves which correspond to the $19.5 \lesssim N \lesssim 23.05$ interval. 
It can be seen from the Fig.\ref{fig:muU}(a) that potential curves have either one or two extrema and tend to zero at $w \rightarrow \infty$. The potentials with two extrema indicate the existence of two stationary solutions. The relative maximum of the potential corresponds to the unstable stationary solution.
The local minimum of the potentials or zero of the $f_{b} (w,m,N )$ corresponds to the equilibrium width $w_{s}$ of the QDs. A small deviation from the equilibrium width leads to the breathing mode oscillations of QD. The frequency of this oscillation is found as the second derivative from the effective potential. 
\begin{equation}
  \Omega^{2}_{0} = \left. {\partial^{2} U(w) \over \partial w^{2} } \right|_{w=w_{s}}
    ={ 5 \Gamma(3M+1) \over 3 \Gamma(5M+1) } \left. {\partial^{2} G \over \partial w^{2} } \right|_{w=w_{s}}.
\label{freq}
\end{equation}
The energy of stationary quantum droplet is 
\begin{equation}
E =4 \pi  \int  \limits_{\ 0}^{\ \ \infty} r^2  { \left[ {1 \over 2} |\psi_r|^{2}
- {\alpha \over 2}|\psi|^{4}
     + {2 \beta \over 5} |\psi|^{5} \, \right] dr }=NG \, .
\end{equation}
The chemical potential is found as 
\begin{equation}
  \mu= \left.{\partial E \over \partial N}\right|_{w=w_{s}} =
    \left. {\partial (NG) \over \partial N}\right|_{w=w_{s}} < 0\,.
\label{mu}
\end{equation}
In Fig.~\ref{fig:muU}(b) we show the dependencies of norm $N$ on chemical potential $\mu$. The red line is found from Eq.~(\ref{mu}), points are found from imaginary time simulations of Eq.~(\ref{gpe})~\citep{Yang2010}.
The value of the derivative $d \mu /dN < 0$ is negative, which indicates the stability of the QDs,
according to Vakhidov-Kolokolov criteria~\citep{Vakhitov1973}.
At large $N$ the density of the QDs reaches a uniform saturation density with low compressibility therefore it develops bulk energy which is typical for flat-top QDs. These values can be estimated by means of the Thomas-Fermi (TF) approximation, see Ref.~\citep{Li2018}.
The largest (peak) density is defined as $n_{max}\equiv |\psi_{max}|^2$. The bulk energy is given by $E_{bulk} \approx S_{area} \, \varepsilon(n_{max})$ where $S_{area} \approx N /n_{max}$ is the area of flat-top QDs and $\varepsilon(n_{max})=-\alpha\, n_{max}^2 \,/2 + 2\, \beta\,n_{max}^{5/2}\, /5$ is energy. The minimization of the bulk energy $d E_{bulk}/d n_{max}=0$ yields $n_{\mathrm{TF}}=5\, \alpha /6\, \beta$ corresponding amplitude $A_{\mathrm{TF}}=n_{\mathrm{TF}}^{1/2}$ and the chemical potential is $\mu_{\mathrm{TF}} =d \varepsilon(n_{max})/d n_{max}= -\alpha\, n_{\mathrm{TF}} + \beta \,n_{\mathrm{TF}}^{3/2}$. The straight grey line in Fig.~\ref{fig:muU}(b) represents $\mu_{\mathrm{TF}}$ for a given $\alpha$ and $\beta$.
%

\begin{figure}[htbp]
  \centerline{ \includegraphics[width=6.5cm]{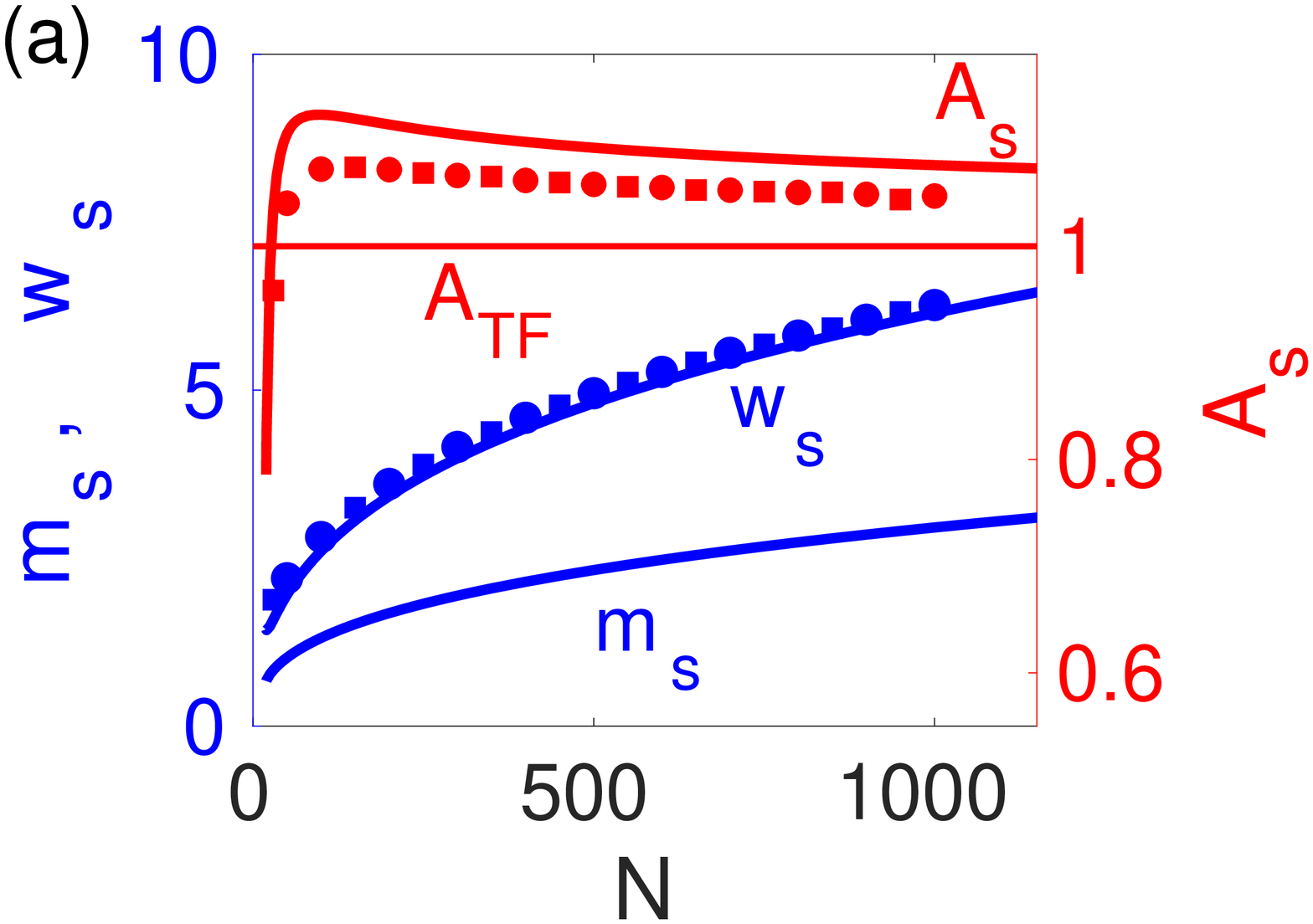} \includegraphics[width=6.5cm]{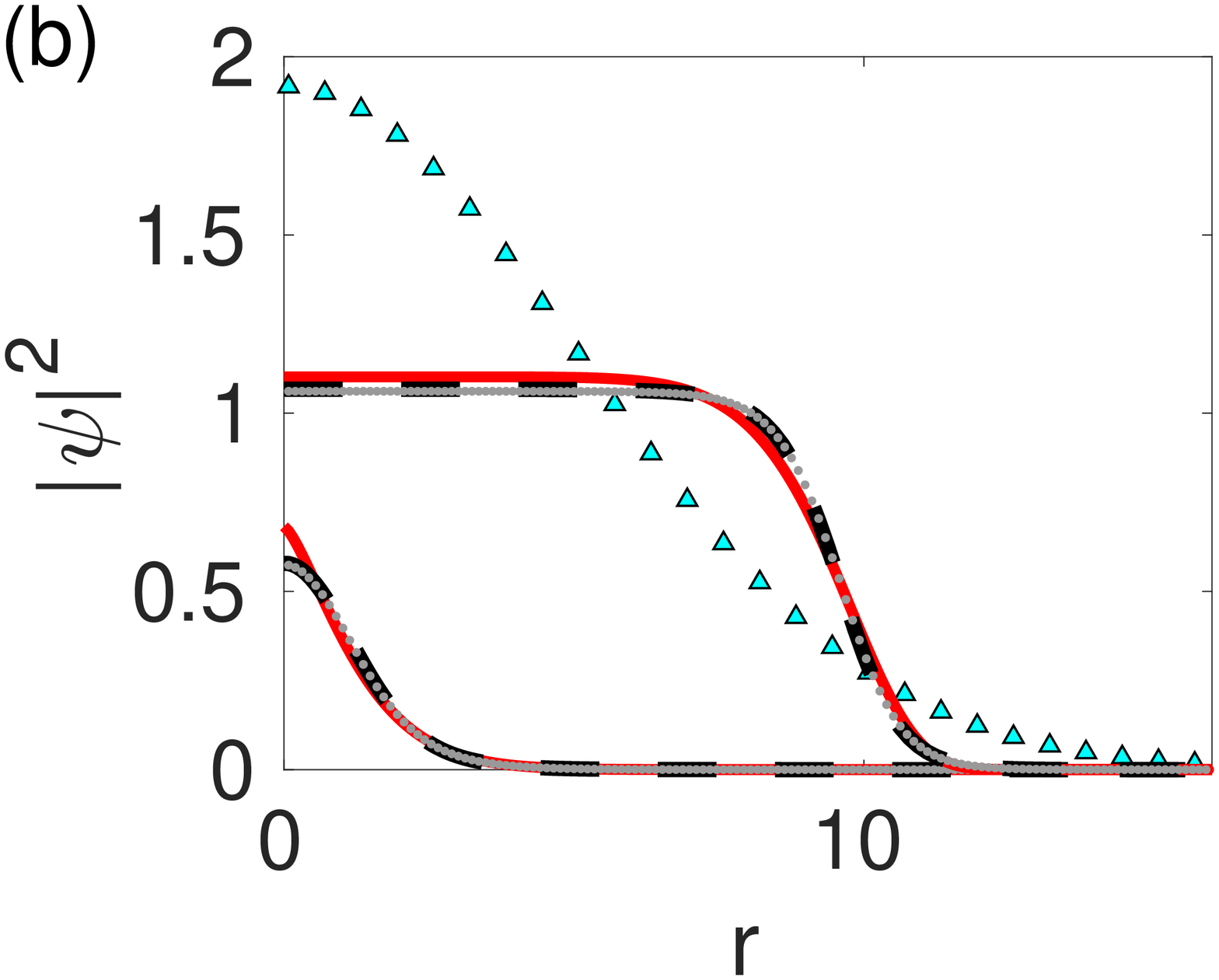}}
\caption{(a) Parameters of stationary QDs, found from the VA (lines) and circular points are found
from imaginary time simulations, and rectangular points are found from numerical simulations
with VA predicted initial parameters after $t=1000$. The red straight line represents the Thomas-Fermi limit. (b) The density profile of QDs. The (red) solid lines are found from Eq.~(\ref{ansatz}) at $t=0$, the (black) dashed lines are found from numerical simulations after adjustment with VA predicted initial condition. The (gray) points are found from imaginary time simulations of Eq.~(\ref{gpe}). The (blue) triangular points are found from VA for $(m,N)=(1,4000)$. The upper curves for $N=4000$, and the bottom curves for $N=20$. }
\label{fig:par}
\end{figure}

The parameters of stationary QDs for different $N$ are presented in Fig.~\ref{fig:par}(a). 
Lines correspond to the prediction of VA. Circular points are parameters of stationary QDs found numerically, using the imaginary time method. 
Figure~\ref{fig:par}(a) shows that the VA predicts well, with accuracy $\sim(1-5)\%$, the stationary parameters in a wide range of $N$.
One can see that the amplitude of QDs tends to the Thomas-Fermi limit for large $N$, while the width increases on $N$. This reflects a liquid-like property of QDs. The stability of the stationary solutions, found from the VA are checked also by numerical modelling of Eq.~(\ref{gpe}). Since the VA deviates from an exact solution, we observe small oscillations of a QD shape. The amplitude of these oscillations decreases on time, and the QD shape tends gradually to the stationary distribution. We call this process an adjustment of QD. In typical simulations, this adjustment ends at $t \approx 200$. We measure the QD parameters at $t=1000$ when the system is in a steady-state, see rectangular points in Fig.~\ref{fig:par}(a), and dashed lines in Fig.~\ref{fig:par}(b). We compared the VA predicted density patterns of QDs with the result of the imaginary time method for bell shape ($N=20$) and flat-top ($N=4000$) profiles in Fig.~\ref{fig:par}(b). The triangular points correspond to the ($m=1$) Gaussian trial function for $N=4000$. One can see that VA based on the Gaussian trial function cannot describe the flat-top states which are characteristic properties of QDs, while the super-Gaussian function gives a good description for this state.

An increase of super-Gaussian parameter $m_s$ on large $N$ means that a density profile of QD approaches the flat-top state. For a given $(\alpha$, $\beta)=(3,5/2)$, we approximate the dependence $m_s$ on $N$ as $m_s(N)=0.275(N-5.371)^{0.344}$. The curve $m_s(N)$ is fitted in the regions $19.5\le N \le 5000$ with following power function $m_s(N)=a_1(N-a_2)^k$.

\begin{figure}[htbp]
\centerline{\includegraphics[width=6.5cm]{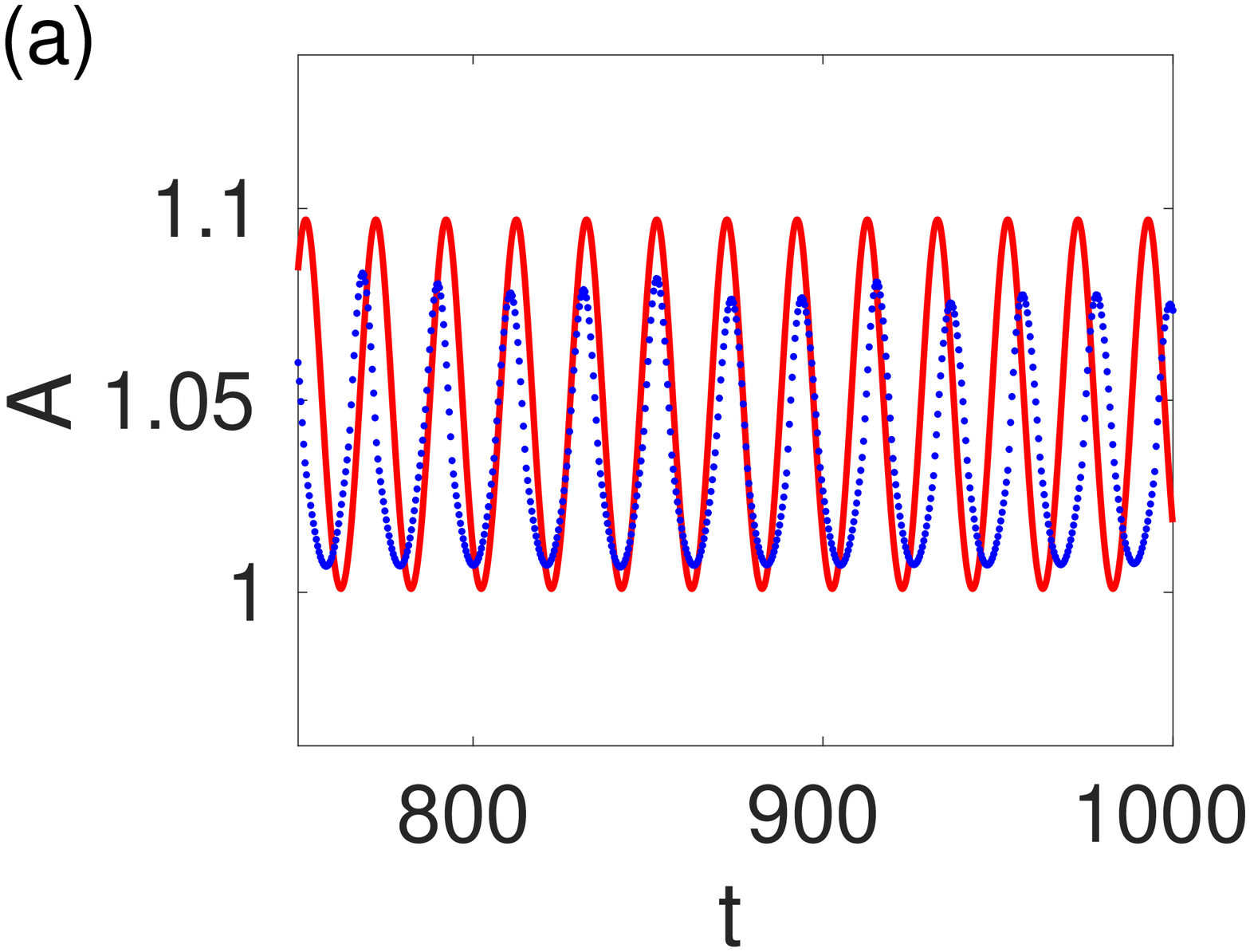} \includegraphics[width=6.5cm]{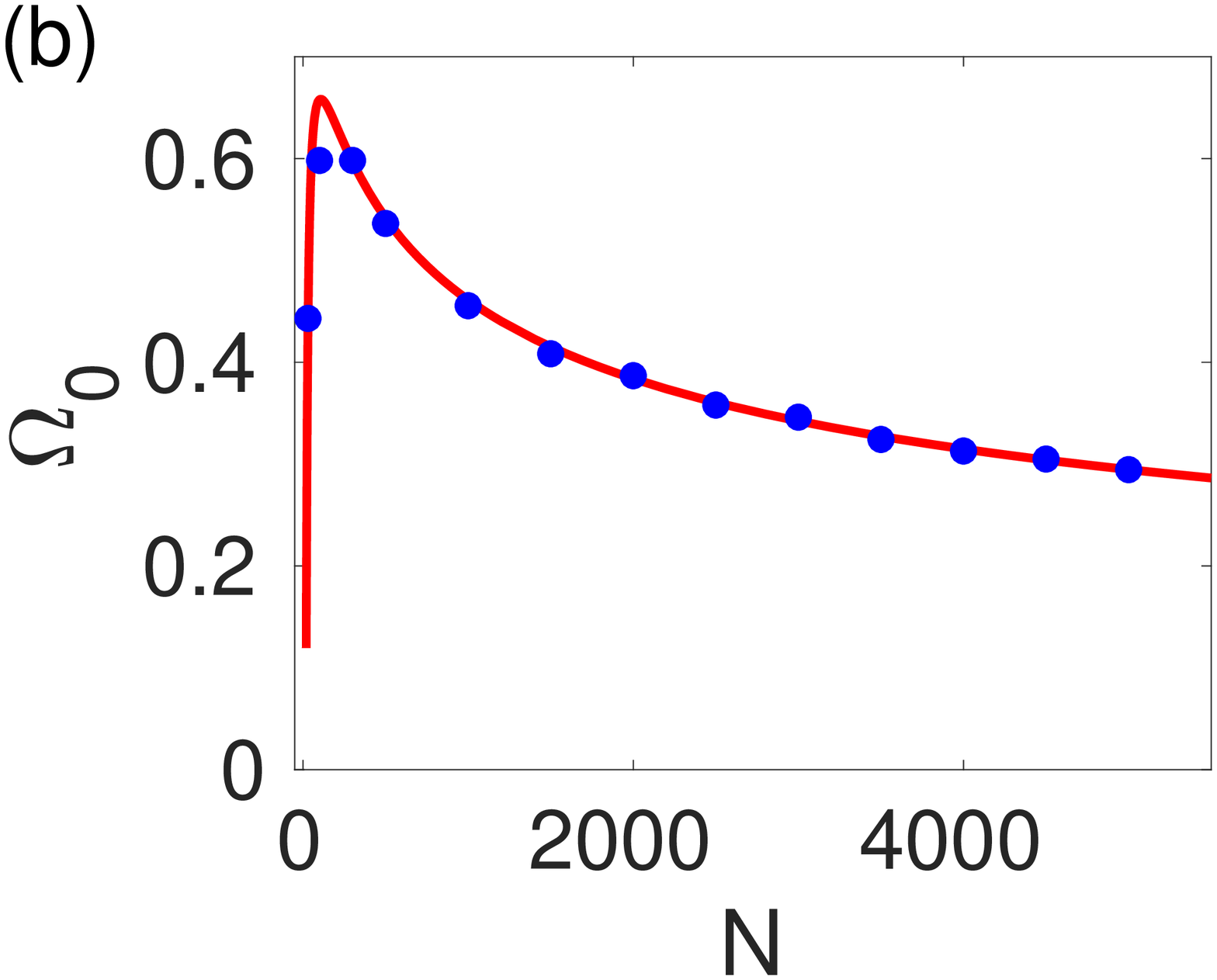}  }
\caption{(a) Oscillation of the QD amplitude found from VA (red solid line) and from numerical simulation of Eq.~(\ref{gpe}) (blue dots) for $N=4000$. (b) The frequency $\Omega_0$ of small oscillations of the parameter A(t) vs norm $N$ for a given $(\alpha, \beta)=(3,5/2)$. The line is found from VA, while points are found from numerical simulations.}
\label{fig:freq}
\end{figure}

In order to find the frequency of the breathing mode oscillations numerically, we use the slightly perturbed initial conditions.
We deviate the VA predicted stationary width $(1 - 10) \%$ and in order to keep the norm unchanged
we also changed the stationary amplitude of the QD see, Eq.~(\ref{norm}).
The typical dynamics of the QD amplitude are demonstrated in Fig.~\ref{fig:freq}(a). 
We see that the VA predicted values of the frequencies are close to the results of numerical simulations. A small phase shift is accumulated during many oscillations. We think that this discrepancy is the result of a small deviation of VA predicted solutions from the ground-state solutions.

Figure~\ref{fig:freq}(b) shows the dependence of the frequency of small oscillations $\Omega_0$ on norm $N$. In numerical simulations, for different values of $N$, the angular frequency is found as $\Omega_{PDE}=2 \pi / \tau$ where $\tau$ is the average period of $A(t)$ oscillation. The average period was found within the t=[200,1000] interval. It is also can be found from the $w(t)$ oscillations. One can see from these figures that, super-Gaussian-based VA gives good descriptions for the dynamics of the QDs.
%

\section{Periodic variation of $\alpha(t)$ and $\beta(t)$}
\label{sec:modulation}
To demonstrate the relevance of VA, we consider a periodic modulation of the parameters $\alpha(t)$ and $\beta(t)$ in time:
\begin{equation}
\alpha=\alpha_0[1+\epsilon_1 \sin(\omega_m t)], \qquad \beta=\beta_0[1+\epsilon_2 \sin(\omega_{m} t+\theta )]
\end{equation}
where $\epsilon_1, \epsilon_2 \ll 1$, $\omega_m$, and $\theta$ are the amplitude and frequency, and initial phase of the modulations, respectively. The parameter $\omega_m$ scaled as $2 \pi\, t_s^{-1}$. In experiments, such modulations can be created by using Feshbach resonance technique~\citep{Chin2010}.

\begin{figure}[htbp]
  \centerline{\includegraphics[width=6.5cm]{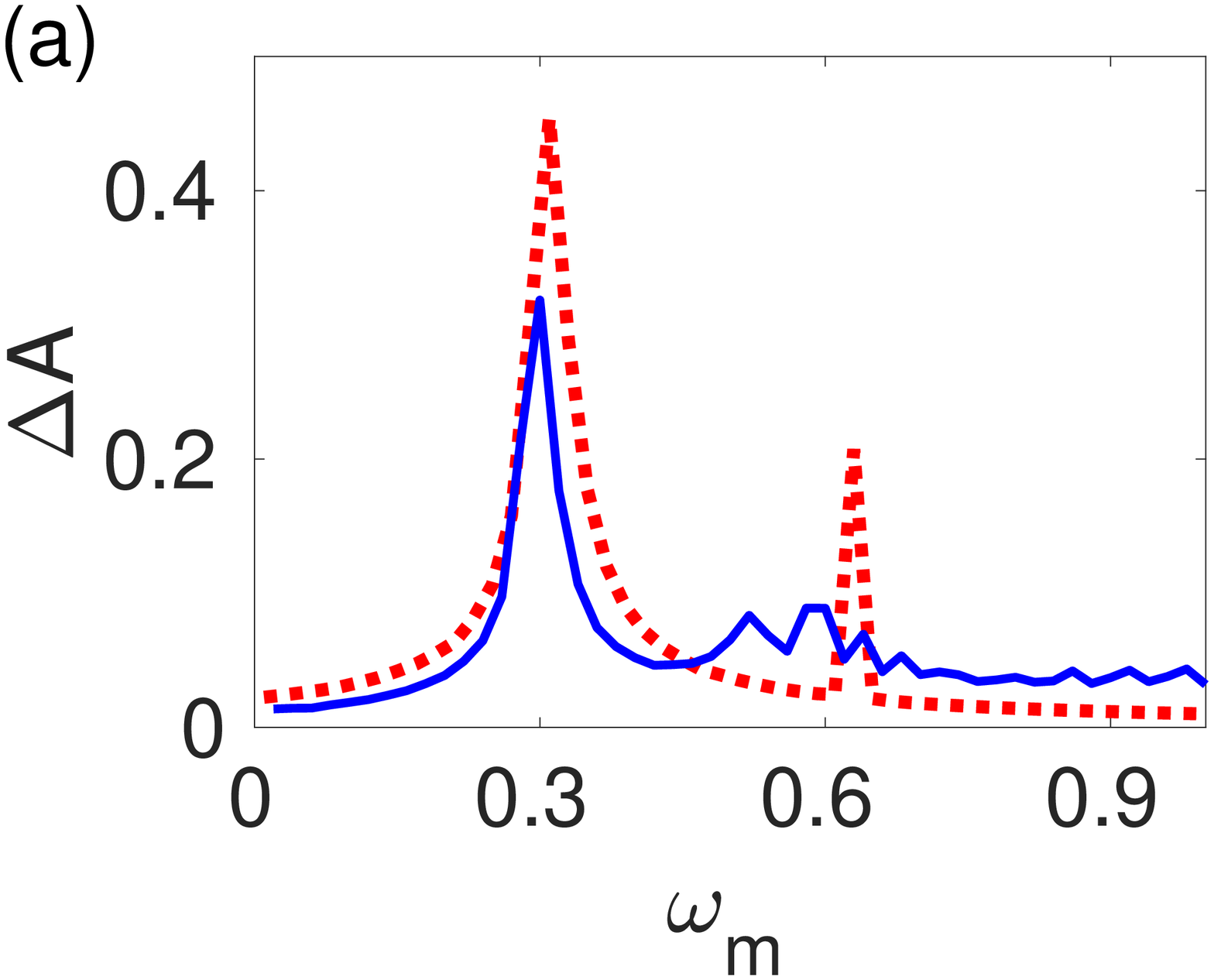} \hskip-0.1cm \includegraphics[width=6.5cm]{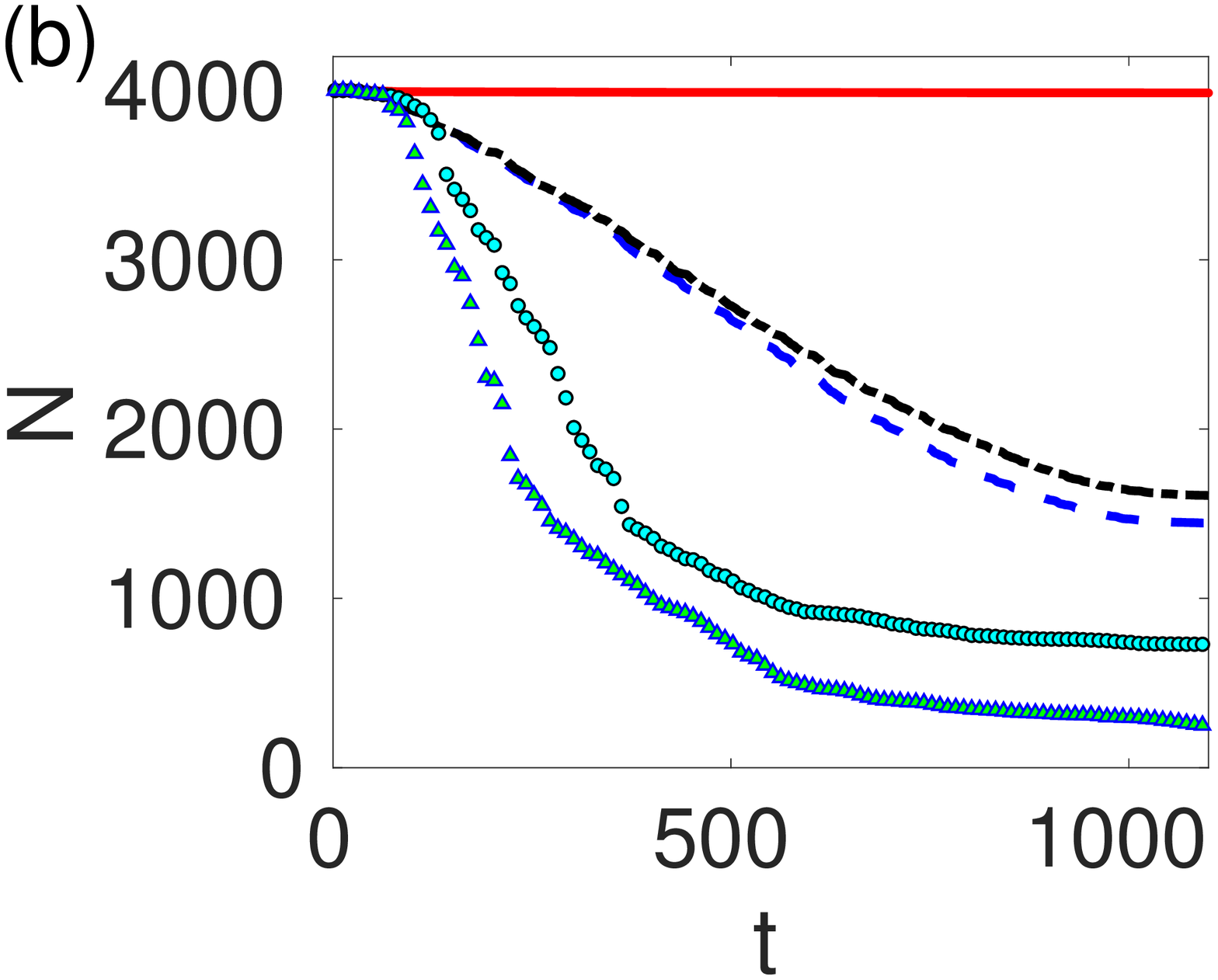}}
\caption{(a) The amplitude difference vs modulation frequency for $\epsilon_1=10^{-2}$ and $\epsilon_2=5 \cdot 10^{-3}$. The line is found from numerical simulations, while points are found from VA. (b) The dynamics of the norm N for different values of modulation amplitude 
for fixed $\omega_m=0.315$. The solid, dashed and dash-dotted lines are for $\epsilon_1=\epsilon_2=0.1$, $(\epsilon_1,\epsilon_2)=(0.1,0.05)$, and $(\epsilon_1,\epsilon_2)=(0.05,0.1)$, respectively. The circular points for $\epsilon_1=\epsilon_2=0.1$ and $\theta=\pi/2$, the triangular points for $\epsilon_1=\epsilon_2=0.1$ and $\theta=\pi$. Other Parameters $(N,\alpha_0,\beta_0)=(4000,3,5/2)$. }
\label{fig:55}
\end{figure}

We include $\theta$ in order to demonstrate the dependence of modulations on the initial phase difference. First we consider $\theta=0$. The periodic variation of $\alpha(t)$ and $\beta(t)$ induces the oscillations of the amplitude and width of the QD. In numerical simulations of Eq.~(\ref{gpe}), we find two different types of QD dynamics depending on the amplitude of the modulations. For small $\epsilon_1$ and $\epsilon_2$, the QD oscillates adiabatically. The amplitude of oscillations increases sharply when the frequency of the modulation $\omega_m$ is close to $\Omega_0$., see Fig.~\ref{fig:55}(a). For $(\alpha_0, \beta_0)=(3,5/2)$, the eigenfrequency $\Omega_0$ is 0.315 while numerical simulation give a value of the resonance frequency $\omega_r=0.30$. This frequency is found in numerical simulations from the dependence of the amplitude difference $\Delta A=A_{max}-A_{min}$ on modulation frequency $\omega_m$, where $A_{max}$ ($A_{min}$) is the maximum (minimum) values of $A(t)$ oscillation on time. 
In the dynamics of QD with small periodic modulations, we observe a beating of QD amplitude and width, which is typical for forced oscillations. During adiabatic oscillations, the number of particles in QD is almost constant, see solid line in Fig.~\ref{fig:55}(b). 
%

The second type of dynamics realized for sufficiently large $\epsilon_1 (=\epsilon_2)$, consists in a gradual decay of a QD. During this process, the QD emits particles in the radial direction. These emissions can be observed as low-density waves, moving away from the QD. We also consider the different values of modulation amplitude, $\epsilon_1 \neq \epsilon_2$. In these cases QD decay faster than $\epsilon_1 = \epsilon_2$, see dashed and dash-dotted lines in Fig.~\ref{fig:55}(b). 

We find that these two forces acting on a QD are antiphase, strongly depend on $\theta$. For $\theta=0$ these two forces almost balance each other, see solid line in Fig.~\ref{fig:55}(b). The fastest decay corresponds to the $\theta=\pi$, see points in Fig.~\ref{fig:55}(b). We see that periodic modulations can be used as an effective tool for changing the thermodynamic states (liquid and vapour) of the BECs also for controlling possible breathing mode oscillations of QD. 

Let us expand the Eq.(\ref{wtt}) by a series of small deviation $\xi$ from the equilibrium width $w_s$, $w=w_s+\xi$, where $\xi \ll w_s$. The dynamics of $\xi$ is described by the following linearized equation:

\begin{equation}
\ddot{\xi}+\left. {\Omega_0^2} \right|_{w=w_{s},\alpha=\alpha_0,\beta=\beta_0}\, \xi =B \sin(\omega_m t) \, ,
\end{equation}
where $\ddot{\xi}$ represents the second derivative of $\xi$ with respect to $t$, and
$$B=-\cfrac{3 c \alpha_0 G_1 \epsilon_1}{w_s^4} + \cfrac{9 c G_2 \beta_0 \epsilon_2}{2 w_s^{11/2}}, 
\qquad c\equiv\cfrac{5\Gamma(3M+1) }{3\Gamma(5M+1)}\,,$$ 
the coefficients $G_1$ and $G_2$ can be found from Eqs.~(\ref{G012}).

\begin{figure}[htbp]
  \centerline{\includegraphics[width=6.5cm]{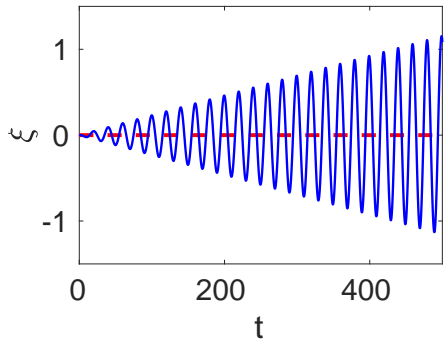} }
\caption{ The dynamics of $\xi$ for different modulation amplitudes. The (red) dashed line for $\epsilon_1=0.1$ and $\epsilon_2=0.102$. The (blue) solid line for $\epsilon_1=\epsilon_2=0.1$. Other parameters $(\alpha_0,\beta_0,N)=(3,5/2,4000).$}
\label{fig:5}
\end{figure}

In the case of symmetric modulation, $\epsilon_1 = \epsilon_2 \ll 1$, the coefficient $B$ is very small $B<10^{-3}$, which means that these two modulations almost compensate for each other. From the condition of $B=0$ we found the following relation $\epsilon_2=2 \alpha_0 G_1 w_s^{3/2} \epsilon_1 / 3 \beta_0 G_2$ between the modulation amplitudes. If $\epsilon_1$ and $\epsilon_2$ fulfils this condition, the modulations totally cancel each other, see dashed line in Fig.~\ref{fig:5}. This condition is valid for any values of modulation frequency $\omega_m$. In this case, the dynamics of QD parameters do not change over time. The accuracy of this relation is confirmed in the dynamics of QD in VA and as well as in numerical simulations. In simulations, to check this relation, we use an imaginary time profile as an initial condition and observe small negligible oscillations of QD parameters around stationary values. 
These linear analyses show, if the modulation amplitudes $\epsilon_1$ and $\epsilon_2$ do not fulfil above condition, dynamics of $\xi$ depend on the modulation frequency $\omega_m$. When the frequency $\omega_m$ is equal to eigenfrequency $\Omega_0$, the system is under resonance oscillations, see solid line in Fig.~\ref{fig:5}. Near resonance frequency, we observe a beating. One can see from Fig.~\ref{fig:5} that in resonance frequency, oscillation amplitude tends to infinity. However, as mentioned above, in nonlinear dynamics with resonance frequency we observe a beating of QD parameters. It is because in a nonlinear system oscillation amplitude and frequency are dependent. 

\section{Numerical simulations and estimation of parameters }
\label{sec:simulation}
In numerical simulations of Eq.~(\ref{gpe}) we use the split-step Crank–Nicolson method with 2048 discrete points and the spatial region of size $L \in [0 - d]$ where $d = 20 - 80$, depending on the QD width, the time step was $dt=10^{-4}$. To prevent reflections of waves emitted by a QD we use the absorbing boundary conditions. For solving the linear part of the system we use the forward and backward sweep method~\citep{Taha1984}.

The width of  the QDs is found numerically by using the mean square radius $<r^2>$:
\begin{equation}
<r^2>= \cfrac{\int \limits_{0}^{\infty }{r^4 |\psi|^2 dr} }{ \int \limits_{0}^{\infty }{r^2 |\psi|^2 dr} }\,.
\label{meanR2}
\end{equation}
Substituting Eq.~(\ref{ansatz}) into Eq.~(\ref{meanR2}) and integrating yields $<r^2>=w^2 \Gamma(5M)/$ $\Gamma(3M)$.
Then, the width of  the QD is found as:
\begin{equation*}
w=\left( \cfrac{\Gamma(3M)}{\Gamma(5M) } <r^2>  \right)^{1/2}.
\label{num_w}
\end{equation*}
The relation between $w$ and full width at half maximum is $w_{\mathrm{FWHM}}=2(\log2)^M w$. 

These quasi-one-dimensional simulations are also supported by real 3D simulations using the split-step Fourier method with $256\times256\times256$ grid points.

By using standard transformation $\psi=\varphi/r$, Eq.~(\ref{gpe}) can be written in following form:
$$
      i \cfrac{\partial \varphi}{\partial t} + \cfrac{1}{2}\, \cfrac{\partial^2 \varphi}{\partial r^2}
       + \cfrac{\alpha}{r^2} \, |\varphi |^{2}\, \varphi - \cfrac{\beta}{r^3}\, |\varphi|^{3}\, \varphi =0 \, .
$$
One solves this equation by using the split-step Fourier method which is much faster than the sweep method. 
Initial condition are taken as $\varphi= r\,\psi$, where $\psi$ is found from Eq.~(\ref{ansatz}) with VA predicted parameters.

Let us estimate the parameters of our model for realistic experiments. We consider $^{39}\mathrm{K}$ atoms in different spin states with mass $m=6.49 \times 10^{-26}$ kg. 
Intra- and inter-species scattering length are $a_{11}=a_{22}=a=50\,a_0$, and $a_{12} =-42 \,a_0$ where $a_0$ is Bohr radius, so that the modified coupling constants satisfies $|\delta g| \ll g$\,. The characteristic scales of our system with $|\alpha|=3$ and $|\beta|=5/2$ are $r_s \approx 0.413$ $\mu m$, $t_s \approx 0.11$ ms, $N_s \sim \psi_s^2 r_s^3 \approx 460$. Dimensionless $t=1000$ corresponds to $105$ ms. In numerical simulations, we take the integration domain as $d=80$, which corresponds to physical units $\approx 33$ $\mu m$. Real atom numbers and size of the QD that correspond to the N=1000 are $4.65\cdot 10^5$ and $3$ $\mu$m, respectively. These parameters are in a range of typical experiments. 

\section{Conclusions}
\label{sec:conc}

We have shown, that similarly to the 1D and 2D cases~\citep{Otajonov2019, Otajonov2020}, the super-Gaussian function is a good approximation for the description of stationary 3D quantum droplets. By using the VA, the dynamical equations have been obtained for the parameters of QD. For large $N$ the width of the QD increases, while the maximum density approaches to a constant (the Thomas-Fermi limit) which shows the behaviour of incompressible liquids. It is also shown that the VA gives a good description of the dynamics of QDs. The frequency of small oscillations of QD shape is found. The periodic modulations of parameters $\alpha$ and $\beta$ with frequency $\omega=\Omega_0$ lead the resonance oscillations. We have observed the different regimes for the dynamics of QDs depending on the amplitude of the external modulations $\epsilon_1$ and $\epsilon_2$, such as adiabatic oscillations for small $\epsilon_1$ and $\epsilon_2$ and decay of QDs for sufficiently large modulation amplitudes. We have shown that modulations of $\alpha$ and $\beta$ are antiphase, and found the relation when these modulations entirely cancel out.

\section*{Acknowledgements}
This work was supported by the Ministry of Innovative Development
of the Republic of Uzbekistan. The author Sh. R. Otajonov thanks 
Prof. F. Kh. Abdullaev and Dr. E. N. Tsoy for their valuable 
comments and discussions.

\end{document}